\documentclass[letterpaper, 10 pt, conference]{ieeeconf}  

\IEEEoverridecommandlockouts                              

\usepackage{graphics} 
\usepackage{epsfig} 
\usepackage{mathptmx} 
\usepackage{times} 
\usepackage{amsmath} 
\usepackage{amssymb}  
\usepackage{bbm}
\usepackage{dsfont}

\usepackage{xcolor}         
\usepackage{soul}           
\usepackage{multirow}
\usepackage{hyperref}
\usepackage{mathrsfs}
\usepackage{svg}

\title{\LARGE \bf
Mixed Traffic Control and Coordination from Pixels
}

\author{Michael Villarreal$^{1}$, Bibek Poudel$^{1}$, Jia Pan$^{2}$, Weizi Li$^{1}$
\thanks{$^{1}$Michael Villarreal, Bibek Poudel, and Weizi Li are with the Min H. Kao Department of Electrical Engineering and Computer Science at the University of Tennessee, Knoxville, TN, USA {\tt\small \{tvillarr, bpoudel3\}@vols.utk.edu; weizili@utk.edu}}%
\thanks{$^{2}$Jia Pan is with the Department of Computer Science at the University of Hong Kong, China {\tt\small jpan@cs.hku.hk}}%
}

\begin{document}

\maketitle
\thispagestyle{empty}
\pagestyle{empty}

\begin{abstract}

Traffic congestion is a persistent problem in our society. 
Previous methods for traffic control have proven futile in alleviating current congestion levels leading researchers to explore ideas with robot vehicles given the increased emergence of vehicles with different levels of autonomy on our roads. 
This gives rise to mixed traffic control, where robot vehicles regulate human-driven vehicles through reinforcement learning (RL). 
However, most existing studies use precise observations that require domain expertise and hand engineering for each road network's observation space. Additionally, precise observations use global information, such as environment outflow, and local information, i.e., vehicle positions and velocities. Obtaining this information requires updating existing road infrastructure with vast sensor environments and communication to potentially unwilling human drivers. 
We consider image observations, a modality that has not been extensively explored for mixed traffic control via RL, as the alternative: 1) images do not require a complete re-imagination of the observation space from environment to environment; 2) images are ubiquitous through satellite imagery, in-car camera systems, and traffic monitoring systems; and 3) images only require communication to equipment. In this work, we show robot vehicles using image observations can achieve competitive performance to using precise information on environments, including ring, figure eight, intersection, merge, and bottleneck.
In certain scenarios, our approach even outperforms using precision observations, e.g., up to $8$\% increase in average vehicle velocity in the merge environment, despite only using local traffic information as opposed to global traffic information.


\end{abstract}

\section{Introduction}


Traffic congestion is a prevalent challenge in modern society, causing delays, gridlocks, and substantial economic losses. Traditional traffic management methods such as traffic lights, stop signs, and ramp meters have proven insufficient in alleviating the current level of congestion~\cite{goodwin2004economic, arnott1994economics}. 
As more vehicles with varying degrees of autonomy are introduced into our transportation system, the idea of mixed traffic control, which involves the use of robot vehicles (RVs) to regulate human-driven vehicles (HVs), is gaining popularity as a potential solution. Studies have shown the effectiveness of this approach in stabilizing traffic on roads of different configurations, including ring and figure-eight roads~\cite{wu2021flow}, merge and bottleneck roads~\cite{vinitsky2018benchmarks}, intersections~\cite{vinitsky2018benchmarks,yan2021reinforcement,wang2023learning,Villarreal2024Eco,Wang2024Privacy}. 
Among various control methods for mixed traffic, reinforcement learning (RL) has emerged as a promising tool, as it can handle the complex behaviors of mixed traffic without using predefined models or heuristics~\cite{chou2022lord}.

\begin{figure*}
    \centering
    \includegraphics[width=0.87\linewidth]{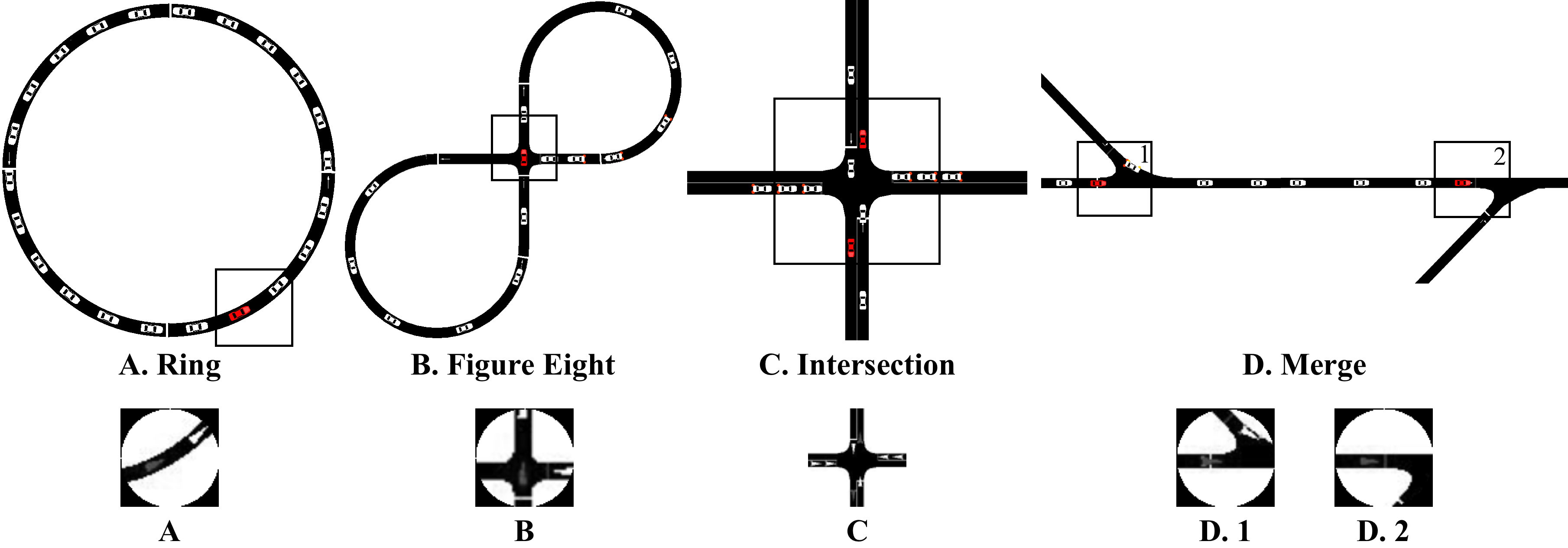}
    \vspace{-1.0em}
    \caption{
    We experiment on five mixed traffic control environments (bottleneck shown in Fig~\ref{fig:hetero_bn}), with image observations presented beneath them. Robot vehicles (RVs) are red, while human-driven vehicles (HVs) are white. With image observations, HVs are cyan to provide contrast from the white background. We use static, grayscale, $84\times84$ images centered over RVs (or intersection) that provide only local information. Merge and bottleneck are multi-agent, while the other three are single agent.
    }
    \label{fig:networks_and_states}
    \vspace{-1.5em}
\end{figure*}

Existing studies of mixed traffic control via RL predominantly uses precise traffic conditions as policy input~\cite{vinitsky2018benchmarks, yan2021reinforcement, wu2017emergent, vinitsky2018lagrangian, kreidieh2018dissipating}: the RV receives both exact global information such as network throughput and travel time as well as exact local information such as nearby vehicles' positions and velocities. 
While effective, \emph{using precise observations necessitates completely re-designing the observation space across different road environments}~\cite{wu2021flow, vinitsky2018benchmarks, yan2021reinforcement, villarreal2023can}, which requires costly hand-engineering and domain expertise.  
For example, the figure-eight environment (Fig.~\ref{fig:networks_and_states}B) uses all vehicles' positions and velocities, while the bottleneck environment (Fig.~\ref{fig:hetero_bn}) uses averaged position and velocity of HVs and RVs in combination with network outflow.  

In practice, for RV to obtain accurate global information, road sensor infrastructure is needed for data collection. 
Overhauling current road networks for this purpose requires substantial expenses. 
To receive precise local information, RV needs to establish vehicle-to-vehicle (V2V) and vehicle-to-infrastructure (V2I) communications.
V2I again needs augmented infrastructure with a multitude of sensors.
V2V, on the other hand, requires HVs to broadcast precise traffic information and engage in constant communication, which is difficult to achieve. 
An alternative to avoid these pitfalls is using image observations (instead of precise observations), a modality commonly seen in robotics research~\cite{sinha2022s4rl, zhu2023viola, shah2023lm, yu2021visual}, but rarely in mixed traffic control.  

In this work, we use bird's-eye view images centered on RV as input to RL policies for mixed traffic control (see Fig.~\ref{fig:networks_and_states}).
The images have generic resolutions and only capture local traffic information.  
Our approach enjoys several benefits.
First, using images as input enables end-to-end training, thus \emph{avoiding the need for manually designing observation spaces}. 
The process of capturing image observations can be repeated over different road environments.
Second, using images can \emph{enable RVs to generalize to new environments as it omits the global information of road networks}. 
This feature is particularly useful since the V2I support (which is required to gain the global information) could vary significantly in different areas. 
Third, HVs are relieved from V2V communication---\emph{a setting greatly enhances the practicality of mixed traffic control}. 
Fourth, imagery about traffic conditions is ubiquitous. 
Satellite imagery can capture traffic in both cities and rural areas, where communication is sparse. 
Our road infrastructure is equipped with ubiquitous camera systems~\cite{caltranspems,coloradotrans}, 
which provide real-time images of traffic. 
Furthermore, modern cars' cameras can capture $360^\circ$ view of the surroundings, which can be used to develop image observations in real-time~\cite{li2022bevformer, xie2022m, zhang2022beverse, huang2023fast}. 
The effectiveness of our approach is demonstrated via comprehensive experiments.   
In summary, our contributions are as follows.  
\begin{itemize}
    \item We use image observations as policy input for RVs in mixed traffic control. 
    These images are both generic and local: they only record the local surroundings of the RVs. In contrast, global information is needed by the RVs for control when using precise observations.
    \item We demonstrate the same-level performance of our approach to using precise observations on various road environments, including a ring road, a figure-eight road, and a merge scene. 
    \item We further achieve improved performance in several cases as compared to using precise observations, e.g., an $8\%$ increase in average vehicle velocity in the merge environment. 
\end{itemize}

To the best of our knowledge, our work is the first to perform extensive experimentation on various road environments to demonstrate the feasibility of using image observations for RL-based mixed traffic control in alleviating traffic congestion.

\section{Related Work}


A significant portion of training RVs via RL with images focuses on individual vehicle driving but not controlling entire traffic~\cite{Shen2022IRL,Poudel2022Micro}.
For example, images are used with vision transformers to learn an effective driving policy~\cite{kargar2021vision}, to train RVs to drive in simulation~\cite{dosovitskiy2017carla, cai2021carl, perez2022deep, zhang2021end}, or to prevent crashes by capturing the RVs' surroundings~\cite{cao2020reinforcement}. 
While these studies apply RL and images to RVs, they do not concentrate on traffic control. 

Wu et al.~\cite{wu2021flow} pioneer mixed traffic control using RL. They show the effectiveness of training an RV on smoothing out stop-and-go waves on a ring road. 
Further tests are conducted by Vinitsky et al.~\cite{vinitsky2018benchmarks} on additional environments, including merge, bottleneck, and intersection scenarios. 
Recently, Wang et al.~\cite{wang2023learning} manage to scale up mixed traffic control to real-world, complex intersections while controlling and coordinating hundreds of vehicles. 
While significant advancements have been made in mixed traffic control, all these studies use precise observations as inputs to the RL policy.
These precise observations include both global and local traffic conditions, such as environment outflow and vehicle position and velocity. 

Our work replaces these precise observations with image observations in the training of RL policies for mixed traffic control. This shift to image observations has several benefits. First, it eliminates the requirement for manual design of the observation space for different traffic scenarios, making it a more flexible solution. Second, it can leverage existing traffic infrastructure as image data is readily available from sources such as satellite imagery, traffic monitoring systems, and vehicle surround-view cameras. 

Prior research use bird's-eye view (BEV) images (which is our image observations) with RVs. 
For example, significant research develop BEV images for 3D object detection and segmentation using modern cars' multi-camera vision systems~\cite{li2022bevformer, xie2022m, zhang2022beverse}. Another work by Huang et al.~\cite{huang2023fast} develops a framework for real-time BEV image perception using onboard vehicle chips. 
This research allows using modern vehicles' camera systems for mixed traffic control at the pace needed to take actions.


Several studies explore mixed traffic control with images. However, there are key differences between their efforts and this project.  
A prior work presents a decentralized method for training RVs with images~\cite{maske2019large}, while another shows human driving can be positively augmented using an RL controller trained on local images~\cite{bayencscrs}. 
Both studies only concern the ring and/or figure eight environments and the focus is not alleviating traffic congestion in varied road environments.

\section{Methodology}
\label{sec:methodology}

We introduce our problem formulation and then outline the details (observation and reward) for our five road environments. 
We also provide details on precise observations~\cite{wu2021flow, vinitsky2018benchmarks} and compare them to image observations. 


\subsection{Preliminaries}
\label{sec:rl_prelims}
We model the RL problem as a Partially Observable Markov Decision Process (POMDP) represented by a tuple ($\mathcal{S}$, $\mathcal{A}$, $\mathcal{P}$, $\mathcal{R}$, $p_0$, $\gamma$, $T$, $\Omega$, $\mathcal{O}$) where: $\mathcal{S}$ is the state space; $\mathcal{A}$ is the action space; $\mathcal{P}(s'|s,a)$ is the transition probability function; $\mathcal{R}$ is the reward function; $p_0$ is the initial state distribution; $\gamma\in(0, 1]$ is the discount factor; $T$ is the episode length (horizon); $\Omega$ is the observation space; and $\mathcal{O}$ is the probability distribution of retrieving an observation $\omega \in \Omega$ from a state $s \in \mathcal{S}$. 
At each timestep $t \in [1,T]$, a robot vehicle (RV) uses its policy $\pi_{\theta}(a_t|s_t)$ to take an action $a_t$ $\in$ $\mathcal{A}$, given the state $s_t$ $\in$ $\mathcal{S}$. The RV's environment provides feedback on action $a_t$ by calculating a reward $r_t$ and transitioning the agent into the next state $s_{t+1}$. The RV's goal is to learn a policy $\pi_{\theta}$ that maximizes the discounted sum of rewards, i.e., return, $R_t = \sum^{T}_{i=t}\gamma^{i-t}r_i$. We use Proximal Policy Optimization~\cite{schulman2017proximal}, a model-free, on-policy algorithm, to learn the optimal policy. 

\subsection{Road Environments}
We train RVs using RL on five road environments (ring, figure eight, intersection, merge, and bottleneck) shown in Fig.~\ref{fig:networks_and_states} and Fig.~\ref{fig:hetero_bn} using image observations. These environments originate from FLOW~\cite{wu2021flow}, a RL framework for traffic management.
We give a brief environment description, compare the differences between image observations and precise observations, and provide the reward functions. 


\subsubsection{Ring}
\label{mpd:ring} 
The ring environment is a widely used benchmark in traffic control~\cite{wu2021flow, chou2022lord}. A single-lane circular road environment consisting of $22$ vehicles with $21$ human-driven vehicles (HVs) and one RV. For $3000$ warmup timesteps, the $22$ vehicles act as HVs. During this warmup, subtle perturbations from imperfect human driving behavior can amplify leading some vehicle standstill. This situation is stop-and-go traffic and acts as a wave backpropagating continually through the ring. After the warmup period, the RV begins taking actions for $3000$ timesteps with the goal to dampen and prevent these waves, thus increasing overall average vehicle velocity. 

Our observation is a gray-scale image of dimensions $84 \times 84$ pixels, centered on a single RV as shown in Fig.~\ref{fig:networks_and_states}. To simulate limited visibility, the image is masked by a circle with a radius corresponding to $28.75$ meters in real world. 
Precise observations are a vector of the RV's velocity, the difference between the leading vehicle's velocity and the RV's velocity, and the difference between the leading vehicle's position and the RV's position. 
This precise observation space has been used to produce state-of-the-art performance~\cite{wu2021flow}.
The action space is the continuous acceleration $[-1,1]$~m/s\textsuperscript{2}. 
The reward function encourages high average velocity and small control actions (acceleration) through a weighted combination:
\begin{equation}
    \text{r} = \frac{1}{n}\sum_{i}v_i - \alpha*\left|a_{RV}\right|,
\end{equation}
where $n=22$, $v$ is vehicle velocity,  $\alpha$ is four (chosen empirically), and $a_{RV}$ is the RV's acceleration. 

\subsubsection{Figure Eight}
\label{mpd:figureeight}
The figure eight environment simulates an intersection in a closed loop with $14$ vehicles--$13$ HVs and one RV. From its shape and the number of vehicles in the environment, queues form among cars trying to cross the intersection. This causes a environment-wide decrease in average vehicle velocity. The RV's objective, over $1500$ timesteps, is to increase all vehicle average velocity.

Our observation is the same as of the ring environment (see Sec.~\ref{mpd:ring}) with an example given in Fig.~\ref{fig:networks_and_states}B. 
The masked circle's dimension corresponds to a $21.25$~m radius. 
Precise observations are the velocities and positions of all vehicles within the figure-eight environment~\cite{wu2021flow}. 
This complete information reflects that the state space is used. 
As our observations are local, i.e., images centered on the RV, learning an optimal policy includes additional, challenging steps of perception and representation learning compared to precise observations.
The action space is the continuous acceleration $[-3,3]$~m/s\textsuperscript{2}. 
The reward function aims to increase all vehicles' velocity in the environment: 
\begin{equation} 
\label{eq:figure_eight}
    \text{r} = \frac{max({\parallel}v_{des}*{\mathds{1}}^{k}\parallel_{2}-{\parallel}v_{des}-V_{\text{all}}\parallel_{2},0)}{{\parallel}v_{des} * {\mathds{1}}^{k}{\parallel}_{2}},
\end{equation}
where $v_{des}$ (desired velocity) is $10$~m/s (chosen empirically) and $k$ is the total number of vehicles in the environment.


\subsubsection{Intersection}
\label{mpd:intersection}
Intersection is an idealized two-way stop where east/westbound traffic flow ($500$ vehicles/hour) is less than north/southbound traffic flow ($1333$ vehicles/hour).
This flow difference causes east/westbound queues as it would otherwise be unsafe to cross the intersection. 
RVs are placed in the north/south directions with a penetration rate of $20\%$. The RVs take actions, over $400$ timesteps, to minimize queue formation and increase average vehicle velocity along the east/west directions. 
This environment allows for studying mixed traffic control in directions absent of RVs since RVs control only the north/south directions. 

Our observation is an $84 \times 84$, grayscale image (shown in Fig.~\ref{fig:networks_and_states}) taken solely at the intersection's center. 
No circle mask is applied to the images.
The image dimensions correspond to $50\text{m}\times 50\text{m}$.
Precise observations include global and local traffic information, and consider a user-defined number of vehicles closest to the intersection~\cite{vinitsky2018benchmarks}. Specifically, precise observations contain each vehicle's velocity, each vehicle's position distance to the intersection, each vehicle's edge number (this identifies if the vehicle is in east/west traffic or in north/south traffic), each edge's density, and each edge's average vehicle velocity. 
Our observation is centered on the intersection for a fair comparison as precise observations collect information using the center as a focal point. However, we envision image observations being collected from the RVs or a fusion between RVs and infrastructure.
Our RVs face a more difficult learning task due to their limited radius observations preventing them from inferring environment-wide information.
Actions taken are defined by the continuous acceleration $[-7,7]$~m/s\textsuperscript{2}. 
The reward function penalizes both vehicle delay and vehicle standstills in traffic: 
\begin{equation}
    \text{r} = -\frac{t * {\sum}((V_{\text{max}} - V_{\text{all}}) / V_{\text{max}})}{\text{n} + \text{eps}} - (\text{gain} * ss_{n}), 
\end{equation}
where $t$ is current timestep, $V_{max}$ is a vector of intersection's speed limit, $V_{all}$ is a all vehicle velocity, $n$ is number of vehicles, 
$eps$ prevents zero division, $gain$ is 0.2, and $ss_{n}$ is the number of standstill vehicles. 
Given the reward is negative, the RVs' goal is to minimize delay and vehicle standstills.

\subsubsection{Merge}
\label{mpd:merge}
The merge environment contains a highway and two merging on-ramps. We expand on the original environment~\cite{vinitsky2018benchmarks}, which only contains the one, right-side on-ramp. The highway and on-ramps respective flows can create stop-and-go waves along the highway and congest the on-ramps, reducing the average velocity and outflow (vehicles/hour). The RVs' goal, for $750$ timesteps, is to minimize such wave formation and increase average vehicle velocity. RVs are only placed on the highway at a $10$\% penetration rate.


Our observation is a stack of five images, each of size $84 \times 84$ (shown in Fig.~\ref{fig:networks_and_states}~D.), centered on the RVs. We observe at most five RVs. If there are less than five RVs present, the remaining stack is padded with black images; if more, the extra RVs are treated as HVs. The image dimensions correspond to $41.25$~m in real world. 
Precise observations are a vector of the following~\cite{vinitsky2018benchmarks}: the velocities of the following and leading vehicle for each RV; the difference in positions between the RV and the following and leading vehicles; and the velocity of each RV. HVs on the on-ramp are observed only if they are following vehicles or have merged onto the highway. 
The action space is the continuous acceleration $[-1.5, 1.5]$~m/s\textsuperscript{2}.
The reward function is: 
\begin{equation}
    \text{r} = \text{Eq.~\ref{eq:figure_eight}} - \alpha \sum_{i \epsilon RVs} \max[h_{\max}-h_{i}(t), 0],
\end{equation}
where $h_{max}$ is empirically set to one and $h_{i}(t)$ is the headway (the time distance between two consecutive vehicles) of an RV at $t$. 
The latter half's objective is to penalize small headways between a RV and a HV to discourage traffic bunching, potentially causing stop-and-go waves.



\subsubsection{Heterogeneous Bottleneck}
\label{mpd:bottleneck}

\begin{figure}
    \centering
    \includegraphics[width=0.9\linewidth]{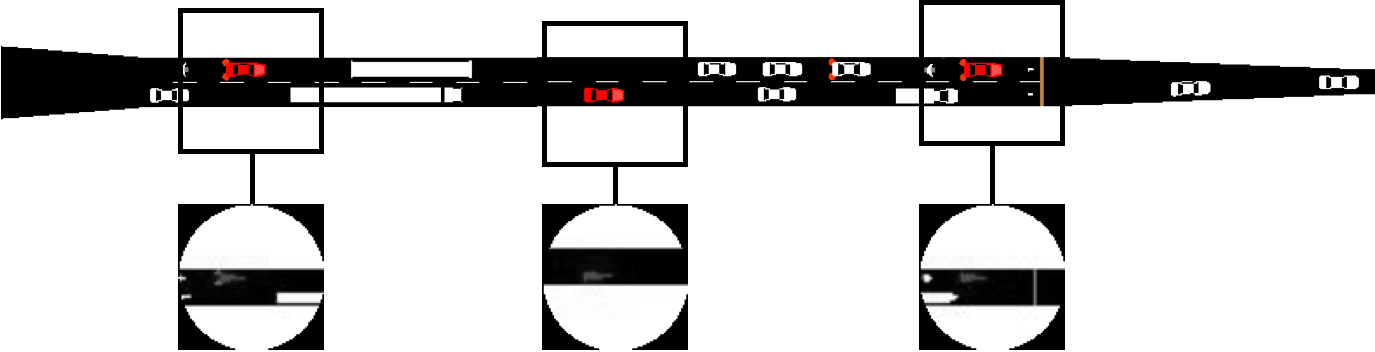}
    \caption{Bottleneck environment with heterogeneous human-driven traffic. We add motorcycles (behind leftmost and rightmost RVs), public buses (in front of leftmost RV), semi-trucks (right of public bus), and delivery trucks (diagonally behind the rightmost RV) alongside regular passenger vehicles.}
    \label{fig:hetero_bn}
    \vspace{-1.5em}
\end{figure}

We also experiment on a heterogeneous bottleneck environment (Fig.~\ref{fig:hetero_bn}). 
The original bottleneck environment~\cite{vinitsky2018benchmarks, vinitsky2018lagrangian} has only four-door passenger vehicles and simulates vehicles experiencing capacity drop~\cite{saberi2013empirical} on a bridge where an environment's outflow significantly decreases after the environment inflow surpasses a threshold. The capacity drop comes from the lanes decreasing from $4 \times l$ to $2 \times l$ to $l$ (where $l$ is a scaling factor and is one for our work). We expand the environment to include heterogeneous HVs comprised (percentage of HVs) of four-door passenger vehicles ($70\%$), semi-trucks ($10\%$), motorcycles ($10\%$), delivery trucks ($5\%$), and public buses ($5\%$) to better reflect bridge traffic. RVs are only four-door passenger vehicles.
The penetration rate of the RV is $10\%$. The RVs' objective is improved outflow in $1000$ timesteps with $40$ prior warmup timesteps.  

Our observation consists of $15$ stacked images, each of size $84 \times 84$ (shown in Fig.~\ref{fig:networks_and_states}), as a maximum of $15$ RVs are placed in the environment. If there are less than 15 RVs, the remaining stack is filled with black images; if more than 15 RVs, additional RVs are treated as HVs. 
The image dimensions correspond to a circle with a radius of $25$~m in real world.
Precise observations (collected on user-defined road segments) contain: mean positions and velocities of HVs, mean positions and velocities of RVs, and environment outflow over the last twenty seconds. 
This observation is difficult to design, consisting of macroscopic and microscopic traffic statistics. 
Global information is considered given segments are examined rather than individual vehicles, which contrasts with out observations containing only local information.
Unlike previously defined environments where the action is an acceleration range, here the action space is the RVs' velocity $[0.01, 23]$~m/s. 
The reward's objective is to increase the outflow  of the environment and reduce the frequency of capacity drops: 
\begin{equation} 
    \text{r} = o_{10}\text{,}
\end{equation}
where $o_{10}$ is outflow over the last $10$ seconds.

\section{Experiments and Results}
\label{sec:results}



\subsection{Experiment Setup}
We train RVs using Proximal Policy Optimization~\cite{schulman2017proximal}, with default hyperparameters from RLlib~\cite{liang2018rllib}.
HVs are operated by Intelligent Driver Model (IDM)~\cite{treiber2000congested} with stochastic noise in the range $[-0.2, 0.2]$ added to account for heterogeneous driving behaviors. 
RVs are trained for $200$ episodes. 

Trained policies are evaluated for $10$ rollouts, and results are presented as averages.
The policies are convolutional neural networks with filters (formatted as [out channels, kernel size, stride]) of $[16, 8, 4]$, $[32, 4, 2]$, and $[256, 11, 1]$ followed by two fully-connected layers. 
Experiments are conducted using i9-$13900$k CPU with $64$GB RAM. 




\subsection{Results}

\subsubsection{Ring}


\begin{figure*}
    \centering
    \includegraphics[width=0.9\linewidth]{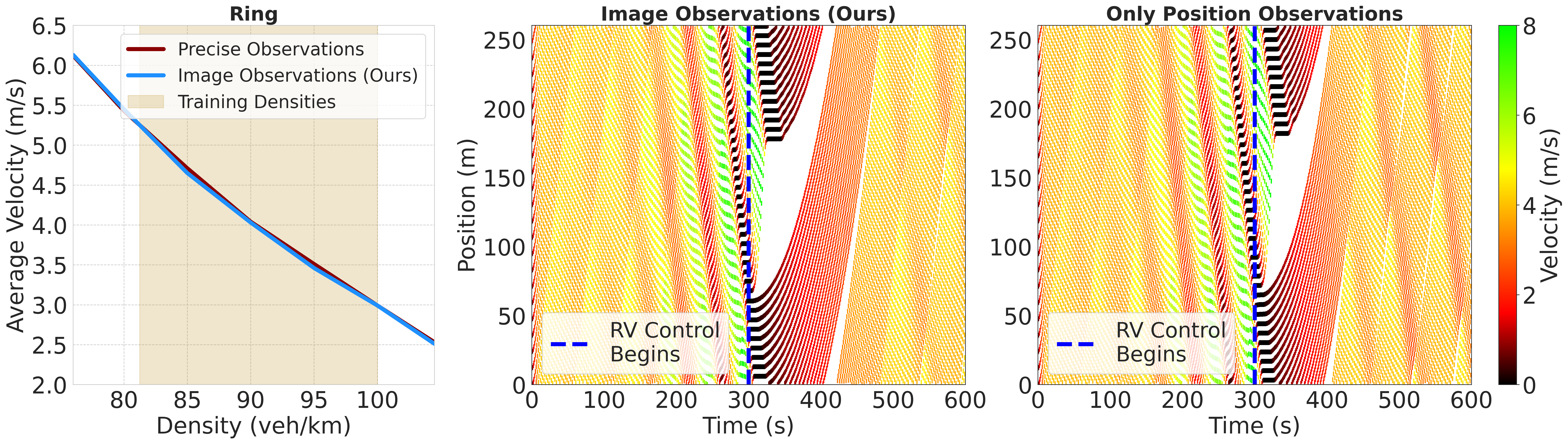}
    \vspace{-1em}
    \caption{LEFT: An RV using image observations prevents stop-and-go waves at all densities, same as an RV using precise observations. MIDDLE and RIGHT: Time-space diagrams showing stop-and-go waves (which form around $200$ to $300$ seconds) being alleviated after RVs start taking action. MIDDLE: An RV trained on image observations prevents stop-and-go waves similar to an RV trained on precise observations. RIGHT: An RV trained using only position information can also prevent stop-and-go waves. This gives further validity of using image observations without explicitly including the velocity information in preventing stop-and-go waves.}
    \label{fig:ring_results}
\end{figure*}

We train the RV on rings with circumference sampled uniformly from $[220, 270]$~m ring length ($[81.25, 100]$ density in Fig.~\ref{fig:ring_results}). For testing, this range is extended to $[210, 290]$~m ring length ($[76, 104.5]$ density in Fig.~\ref{fig:ring_results}).
The results are shown in Fig.~\ref{fig:ring_results} LEFT. 
RVs trained using image observations (blue) prevents stop-and-go waves and achieve the same-level performance as RVs trained using precise observations (red) at all densities.



Fig.~\ref{fig:ring_results} MIDDLE shows the time-space diagram of all vehicles over an episode. 
The shockwaves from $200$ to $300$ seconds (one second equals $10$ timesteps) reflect stop-and-go traffic. 
After $300$ seconds, the image-trained RV takes actions by briefly accelerating and then stabilizes the traffic.

\subsubsection{Figure Eight}

\begin{figure*}
    \centering
    \includegraphics[width=0.9\linewidth]{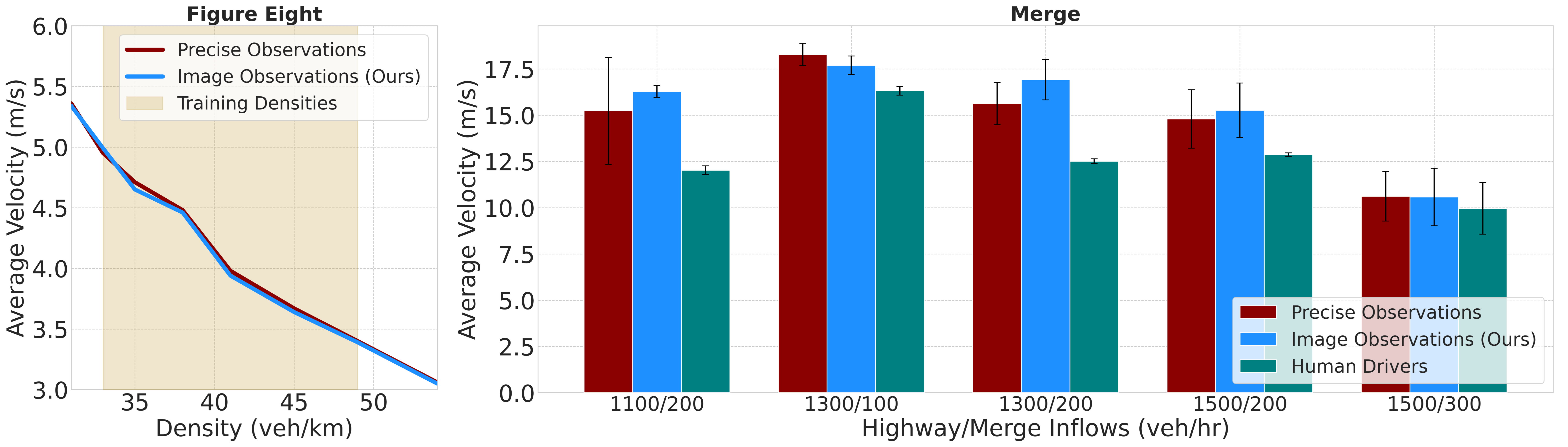}
    \vspace{-1em}
    \caption{LEFT: An RV using image observations achieves mixed traffic control comparable to an RV with precise observations in figure eight.
    RIGHT: Overall, RVs with image observations outperform RVs with precise observations by outperforming RVs with precise observations in $1100/200$, $1300/200$, and $1500/200$ by up to 8\%.}
    \label{fig:three_results}
    \vspace{-1.0em}
\end{figure*}

In prior work~\cite{wu2021flow, vinitsky2018benchmarks}, the RV is trained only on a single, inner-loop radius~\cite{wu2021flow} (inner-loop radius is used to calculate the overall environment length).
We expand the scenario by training on the range $[20,30]$~m ($[33, 49]$ density in Fig.~\ref{fig:three_results} LEFT) and expand this range to $[18,32]$~m ($[31, 54]$ density in Fig.~\ref{fig:three_results} LEFT) during evaluation. The efficacy of a trained RV is measured by the average vehicle velocity at a particular traffic density. 

Fig.~\ref{fig:three_results} LEFT shows that an image-trained RV can achieve the same level of mixed traffic control as a precise-trained RV. This same-level performance is in spite of the image-trained RV only receiving local information compared to the precise-trained RV having complete global/state information.

\subsubsection{Intersection}


\begin{figure}
    \centering
    \includegraphics[width=0.85\linewidth]{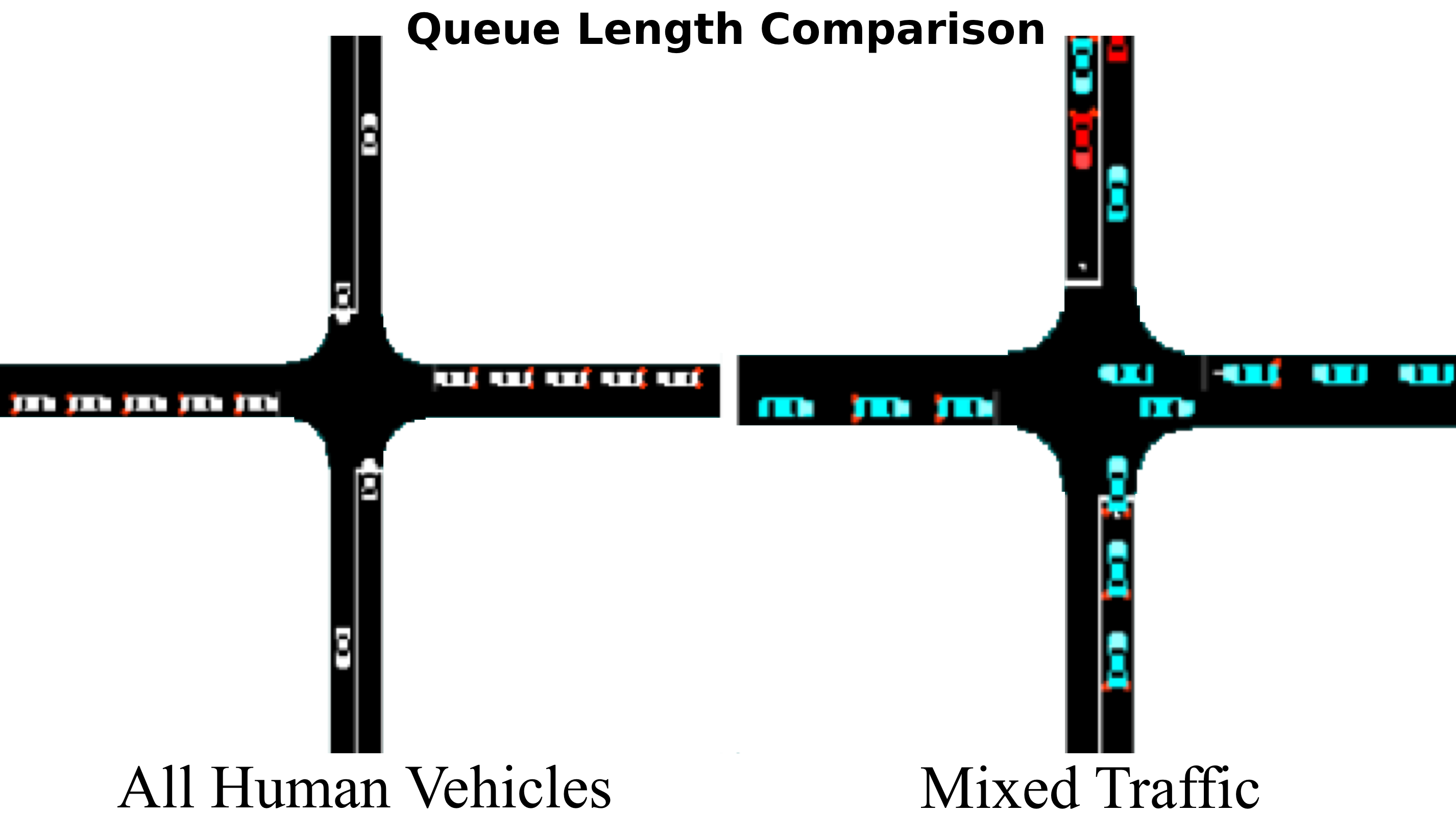}
    \vspace{-1.0em}
    \caption{Comparison between the queue lengths at the end of an episode between all human drivers (cyan in mixed traffic) and mixed traffic using image observations. RVs trained with image observations lessen east/westbound congestion by decreasing queue lengths by two vehicles.}
    \label{fig:ql_results}
    \vspace{-2.0em}
\end{figure}

We consider the average all vehicle velocity and east/westbound queue lengths. RVs with image observations attain $4.75 \pm 0.02$~m/s average vehicle velocity with a three vehicle queue length. RVs with precise observations obtain $5.90 \pm 0.23$~m/s average vehicle velocity with a three vehicle queue length.
RVs with image observations can provide similar performance to RVs with precise observations in regard to queue length. 
The average vehicle velocity of RVs with image observations is less than RVs with precise observations. We believe this performance difference is due to precise global observations knowing exactly what edges vehicles are on and their corresponding velocities, which allow the RVs to know when HVs are at standstill in the east/west directions. 
Both RV types outperform HVs ($3.50 \pm 0.00$~m/s average velocity; five vehicle queue length) in both evaluation metrics.

Fig.~\ref{fig:ql_results} illustrates the development of queues with only HVs versus RVs with image observations. The southbound RV in the right image is slowing down momentarily, which allows east/westbound HVs to safely cross the intersection. Only HVs travel at velocities that cause queue development in the east/west directions.

\subsubsection{Merge}
Fig.~\ref{fig:three_results} RIGHT presents our results. 
We evaluate merge using five combinations of highway/merge inflow rates (x-axis), $\{1100/200$, $1300/100$, $1300/200$, $1500/200$, $1500/300\}$. Merging on-ramps share inflow rates, and we compare average vehicle velocities (y-axis) at those inflows. 
This particular network setup and different inflow combinations that cause varying congestion levels have not been previously studied.

RVs with image observations outperform RVs with precise observations at $1100/200$, $1300/200$, and $1500/200$ inflow rates by $7\%$, $8\%$, and $3\%$, respectively. 
RVs with precise observations outperform RVs with image observations in the $1300/100$ inflows scenario, while both of them have similar performance in the $1500/300$ inflows scenario. 
RVs with image observations provide the largest performance improvement with a $1.29$ m/s--an $8\%$ increase--over using precise observations in the $1300/200$ scenario. 
The $1500/300$ scenario is difficult to learn on (evidenced by both RV types improving performance over HVs the least) as the inflow rates cause sufficient congestion inhibiting the RVs' potential to increase traffic flow from taking intelligent actions.
Overall, RVs with image observations outperform RVs with precise observations.

\subsubsection{Heterogeneous Bottleneck}
We train on two inflows, \{$2300$, $2500$\}, and compare outflow over the last $500$ seconds. The different inflows allow for capturing different congestion levels that allow for improvement through mixed traffic control. At $2300$ and $2500$ inflows, RVs with image observations obtain $1497.60 \pm 26.94$ and $1506.96 \pm 29.17$ outflows, respectively, while RVs with precise observations obtain $1528.56 \pm 49.26$ and $1513.44 \pm 24.90$ outflows, respectively. Both outperform HVs at $2300$ and $2500$ inflows, which achieves $1448.64 \pm 23.40$ and $1447.20 \pm 14.04$ outflows, respectively.
RVs with precise observations outperform RVs with image observations at both inflow rates. Although at $2500$ inflow, RVs with image observations achieve an outflow close to RVs with precise observations. We hypothesize that RVs with precise observations outperform RVs with image observations because precise observations contain network-wide traffic information, while image observations only contain local traffic state information.



\subsubsection{Only Position Observations}
\label{sec:onlypos}
We conduct an additional experiment to test position-only observations in training RVs using precise information in the ring environment. 
The purpose is to analyze whether RVs can still be leveraged to alleviate traffic congestion given only static positional information, similar to positional inference using image observations. 
Precise observations change to a vector of the difference between the RV's position and the leading vehicle's position. 
Fig.~\ref{fig:ring_results} RIGHT shows the time-space diagram for this experiment. A RV with only position information can achieve the same level of performance as of using complete information (i.e., both position and velocity). 
This result solidifies our approach of using image observations without explicitly including the velocity information in preventing stop-and-go wave formation. 

\subsection{Limitations and Discussion}

In this project, we do not assume our RVs to be fully autonomous vehicles with equipment and sensors to allow for complete control. Our RVs control only their acceleration (or velocity), which can be achieved by controlling the throttle signal through a control mechanism using images that do not require significant computational resources to process and receive actions from. This partial autonomy mechanism allows for a human, in a real-world setting, to still control other vehicular functions (such as changing lanes, turning, handling emergency situations, etc), while improving and coordinating traffic conditions. The signals being sent to the vehicle to control traffic conditions can be overwritten by the human driver, which allows for a certain level of safety within the system.

Transmitting image data in a V2V format is comparatively expensive to transmitting precise observations as images have higher dimensions. Drops/delays, resulting in data loss, when communicating with other vehicles is an inevitability~\cite{al2020efficient, aziz2013energy}.
However, this issue can be mitigated by using existing image compression/decompression techniques, allowing significant image dimension reduction~\cite{sonal2007study, choi2019variable, zhou2018variational}. 
Our approach communicates with less vehicles, making the process easier, than precise observations as HVs are not communicated with.
Additionally, our approach requires vehicles/infrastructure to be equipped with image sensors for proper implementation. While some vehicles/infrastructure may be too old or costly, advancements in image sensor technology within cars and transportation infrastructure have increased their prevalence and cost effectiveness.

Despite the generalizability of image observations, the reward functions used are still environment and task specific. A general purpose reward function for transportation environments is still an open problem given how task-specific environments can be. Thus, finding a general purpose reward function is out of scope; however, we believe finding such a general reward function is interesting to pursue in the future to further increase generalization. 



\section{Conclusion and Future Work} 
\label{sec:conclusion}

In this work, we demonstrate the ability of robot vehicles (RVs) to perform mixed traffic control using reinforcement learning (RL) policies trained on image observations. We examine RVs trained on image observations in the ring, figure eight, intersection, merge, and bottleneck environments. Additionally, we expand on the figure eight network lengths trained on, expand the merge environment and inflows trained on, and expand the bottleneck environment to include heterogeneous traffic and inflows trained on. 
We show that RVs trained on image observations have competitive performances to RVs trained on precise observations.

In the future, we aim to advance this study in several directions. 
First, we want to test our approach on more road networks, together with large-scale, long-term traffic simulations~\cite{Li2017CityFlowRecon,Guo2023Long}. This could involve combining multiple road networks together, where RVs must learn to perform multiple tasks concurrently. While simulating these scenarios is feasible, the increased task complexity presents a challenge to the RV's learning process. 
Second, we would like to incorporate additional generic information such as traffic state predictions~\cite{Lin2019Compress} and vehicle trajectories into our observation space for potential improvement. 
Lastly, we want to explore the resilience aspect by taking adversarial attacks and image perturbations into account~\cite{Shen2021Corruption}. 
\section*{ACKNOWLEDGMENT}
This research is supported by NSF IIS-2153426. The authors would also like to thank NVIDIA and the Center for Transportation Research (CTR) at the University of Tennessee, Knoxville for their support.

\bibliographystyle{IEEEtran}
\bibliography{references}

\end{document}